\documentclass[runningheads]{llncs}
\usepackage[T1]{fontenc}
\usepackage{graphicx}

\usepackage{subcaption}
\usepackage{dcolumn}% Align table columns on decimal point
\usepackage{bm,bbm}% bold math
\usepackage{xcolor}
\usepackage{xfrac}
\usepackage{amsmath,amssymb}
\usepackage{pifont}
\usepackage{array}
\usepackage[symbol]{footmisc}
\usepackage{cite}

\usepackage{tabularray}

\usepackage[colorlinks,linkcolor={magenta},citecolor={magenta}, urlcolor={magenta}]{hyperref}

\begin{document}
%
% \title{Medial Surfaces Between Fibers, from Clasps to Knits}
\title{Medial Model of Elastic Curve Contacts, from Clasps to Knits}

\author{Kathryn~Chain\inst{1}\orcidID{0009-0000-7925-8538} \and
Haoye~Sun\inst{1}\orcidID{0009-0004-0630-0183} \and
% Aidan~Foulk\inst{1}\orcidID{...} \and
Michael~S.~Dimitriyev\inst{1,\ast}\orcidID{0000-0001-6384-3644}}
\authorrunning{K.~Chain, et al.}
\institute{Department of Materials Science and Engineering, Texas A\&M University, College Station, TX 77843, USA\\
\textsuperscript{$\ast$}\email{msdim@tamu.edu}
}
\maketitle              % typeset the header of the contribution
\begin{abstract}
Materials constructed from interlocking and tangled fibers are challenging to model and understand due in part to the complex interplay between fiber shape and contact mechanics.
Much of our intuition about contact mechanics comes from the Hertzian description, which is based on interactions between convex bodies.
However, fibers often form clasp shapes, which involve saddle-shaped contact surfaces; the ability for two interacting bodies to conform to each other along a saddle surface confounds many existing mechanical contact models.
Following an observation that contact surfaces between fibers tend to look like minimal, constant mean-curvature surfaces, we seek to test this hypothesis and provide a direct link to geometric modeling methods.
To do this, we leverage the medial axis transformation, which generalizes the notion of a Voronoi tessellation to generating curves and surfaces.
Through computational studies on an orthogonal clasp and a simple knit structure, we find that the medial axis transformation provides a reasonable model for contact surfaces between identical curves.
We also show that a generalization of the medial axis transformation based on radical Voronoi tessellations can be used to well-approximate contact surfaces between non-identical curves.
We apply this ``radical axis transform'' to curves of different radii and demonstrate a trend for surfaces to adopt finite mean curvature.
Our results may point the way towards more sophisticated geometric methods for modeling tangles.

\keywords{Fiber mechanics  \and computational geometry \and tangled metamaterials.}
\end{abstract}
\section{Introduction}

Of the variety of textile structures that have been created throughout human history, knits are instantly recognizable.
Unlike woven fabrics, which are constructed from multiple threads that are made to cross orthogonally in a rectilinear mesh, the basic knit is constructed from a single yarn that traverses each row, or \textit{course}, in a serpentine manner, linking with itself across neighboring rows through the use of loops.
The slipknot-like structure of these loops give rise to fabric that has highly compliant elastic deformations, both in-plane~\cite{singal_programming_2024} and out-of-plane~\cite{Niu2025}.
While the bending rigidity and thread (or yarn) shape plays an important role in determining much of the elastic response of knits, contact interactions are responsible for stress transfer across the material.
However, little is known about contact mechanics for slipknot structures, let alone entire knitted stitches.
This is in large part due to the concavity of inter-thread contact surfaces, inherited from the locally-helical shapes of interacting threads.

Much of what we know about deformations of elastic bodies in contact is derived from the Hertz theory, where convex bodies are brought into contact at either a point (in the case between ellipsoids) or a line (between parallel cylinders)~\cite{Johnson1982,LandauLifshitz1986}.
Using linear continuum elasticity theory, it can then be shown that the force $f$ needed to deform two ellipsoidal bodies into contact scales as $f \sim \delta^{3/2}$ for a sufficiently small indentation depth $\delta$.
This fractional departure from the usual Hooke's law ($f \sim \delta$) arises from a combination of a parabolic strain profile within the material with the effect of increasing the contact area between the two bodies with increasing $\delta$.
While this is generally the case for convex bodies in contact, two non-convex bodies have the potential to conform to each other, resulting in closely matching surface geometries.
The mechanical description of such ``clasps'' necessarily falls outside of the classical Hertz theory but is needed to describe everything from simple capstans~\cite{JUNG2008} to knots~\cite{patil_topological_2020}.
Beyond simple clasps, there is recent evidence that the helical clasps in knitted fabrics are responsible for multi-stable states of the fabric~\cite{Poincloux2018,Crassous2024} and, in the case of tightly-confined fibers, can lead to effectively ``jammed'' states~\cite{gonzalez_pulling_2025}.

In this work, we consider an approximation to the contact surfaces in simple clasps and knitted fabrics motivated by an observation that such contact surfaces are saddle-shaped and may even be modeled as minimal surfaces~\cite{Knittel2020,Wadekar2020_helicoidal,Wadekar2021,Niu2024}.
While liquid interfaces and amphiphiles~\cite{ERIKSSON1994,Deserno2015,Hyde1996,Hyde2001,Canham1970} are known to assemble minimal surfaces in order to minimize a surface free energy cost that scales with area $A$ of the surface, there is a less obvious connection between minimal surfaces and contact geometry.
One can conceive an approximation of the contact energy of the form $U_{\rm int} \approx \int{\rm d}A\, u(\mathbf{x}) = A \overline{u}$, where $\overline{u}$ is an average deformation energy density for two bodies in contact, such that changes in the area $A$ do not significantly alter the average energy density $\overline{u}$.
Recent experimental and finite element studies~\cite{Grandgeorge2021} have demonstrated that the pressure distribution in a pair of clasping fibers can even have multiple local maxima appearing near the periphery of the contact surface rather than its center, suggesting a level of inhomogeneity that is incompatible with a simple area scaling for interaction energy.
However, it is worth considering that such inhomogeneities might be particularly pronounced for incompressible elastomers characterized by neo-Hookean elasticity, whereas spun yarn is highly compressible and may therefore be able to better relax sharp gradients in stress.
Fiber flexibility may also play a role in reducing stress inhomogeneities, promoting conditions under which fibers can conform to each other; in the case of perfectly conforming fibers, we expect that the contact energy is extensive in the contact area.
To examine this hypothesis, we turn to a ``medial surface'' approximation of fiber contact geometry, extracting mean and Gaussian curvature distributions. 
We furthermore consider the cases of differing fiber radii and stiffnesses, which requires a generalization of the medial surface construction based on radical Voronoi tessellations.
Our method creates an approximate contact surface that combines effects of fiber flexibility with a simplified contact mechanics model, facilitating an exploration of different effects.

\section{A 3D elastica model with power-law contact interactions}

\begin{figure}
    \centering
    \includegraphics[width=\linewidth]{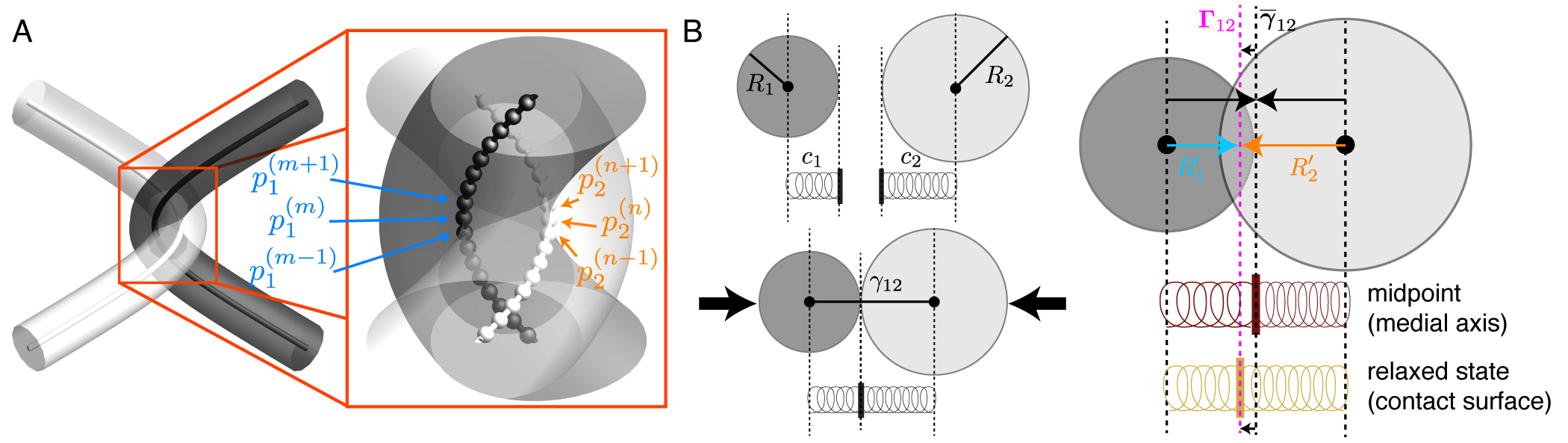}
    \caption{(A) Depiction of the discrete elastic rod approximation of continuous curves, where each curve is approximated as a string of vertices joined by edges. (B) Construction of the interaction model, starting on the top left panel, where two disks representing the cross-sections of the elastic curves are brought into contact (bottom left). By treating each cross-section as an effective nonlinear spring, we can determine the true contact point, which is in general different than the midpoint between the centers of the two cross-sections (right panel).}
    \label{fig:1}
\end{figure}

The mechanics of tangled fibers have been modeled using a variety of approximations, from finite element analysis based in solid continuum mechanics~\cite{Xiong2018,AHN2017,Wadekar2020,Liu2019} to quasi-analytic geometric models~\cite{Postle2002,Choi2006,Poincloux2018PRX}.
In an effort to balance the simplicity and generality of geometric models against the detail and specificity of solid continuum mechanics, we turn to a modified 3D elastica-based yarn model~\cite{Kaldor2008,singal_programming_2024,ding_unravelling_2024}, combining slender-body mechanics with an approximate treatment of contact interactions.
Within this model, degrees of freedom are encoded in a set of arclength-parametrized curves $\left\{\bm{\gamma}_i(s)\right\}$ with short-range repulsive interactions within a given radius around the centerline, as illustrated in Fig.~\ref{fig:1}A.
The energy of an elastic curve is given by
\begin{equation}
    E = \sum_i\int {\rm d}s\,\left[\frac{B_i}{2}\kappa_i^2(s) + \frac{Y_i}{2}\left(\lambda_i^2 - 1\right)\right] + \frac{1}{2}\sum_{i,j}\iint{\rm d}s\,{\rm d}s'\, u_{ij}(s,s')
\end{equation}
where $B_i$ is the bending modulus of curve $i$, $Y_i$ is its Young's modulus, and $\kappa_i$ and $\lambda_i$ are, respectively, the curvature and stretch at arclength parameter $s$.

While prior computational models of soft fiber contact based on energy minimization---particularly those developed for the simulation of knitted fabrics~\cite{singal_programming_2024,ding_unravelling_2024}---essentially considered repulsions between the centerlines of the fibers without any \textit{explicit} deformation of the cross-section, we adopt a different approach.
Here, we consider the contact surface between curve segments as a separate degree of freedom that can be freely varied to minimize the total elastic energy, much like the centerline degrees of freedom; the situation is illustrated in Fig.~\ref{fig:1}B.
In this generalized approach, we consider an interaction potential of the form
\begin{equation}\label{eq:contact_int}
    U_{{\rm int},ij} = \int{\rm d}s\int{\rm d}s' u_{ij}(s,s') = \int{\rm d}s\int{\rm d}s'\left[u_{i}\left(\delta_i\right) + u_{j}\left(\delta_j\right) \right]
\end{equation}
where $u_i$ is the energy density representing the cost of deforming fiber $i$ by displacements $\delta_i$.
A contact interaction locally deforms a fiber radius from $R_i$ to $R'_i = R_i - \delta_i$.
When in contact, the distance $\gamma_{ij}(s,s') \equiv  \lVert \bm{\gamma}_j(s') - \bm{\gamma}_i(s)\rVert$ between curves $i$ and $j$ is simply the sum of these deformed radii, namely $\gamma_{ij} = R'_i + R'_j$.
Therefore, we are left with the relationship between cross-sectional deformations,
\begin{equation}
    \delta_1 + \delta_2 = \overline{\delta} = R_i + R_j - \gamma_{ij}
\end{equation}
where we define $\overline{\delta}_{ij} \equiv \delta_i + \delta_j$ along with $\eta_{ij} \equiv \delta_i - \delta_j$.
The contact point $\bm{\Gamma}$ is then located at
\begin{equation}
\bm{\Gamma}_{ij}(s,s') = \overline{\bm{\gamma}}_{ij}(s,s') + \frac{1}{2}\hat{\bm{\gamma}}_{ij}(s,s')\left(R_i - R_j - \eta_{ij}(s,s')\right) \,,
\end{equation}
where the first term is the midpoint $\overline{\bm{\gamma}}_{ij}(s,s') \equiv (\bm{\gamma}_i(s) + \bm{\gamma}_j(s'))/2$ between the two curve segments and the second term is a correction due to asymmetry in cross-section radius or stiffness and points in the direction of the vector $\bm{\gamma}_{ij}(s,s') \equiv \bm{\gamma}_j(s') - \bm{\gamma}_i(s)$.

The integrand of Eq.~\ref{eq:contact_int} is minimized with respect to variations in $\eta_{ij}$ when
\begin{equation}\label{eq:contact_int_equilibrium_conditions}
    \frac{{\rm d}u_i}{{\rm d}\delta_i}\Bigg\vert_{\delta_i = (\overline{\delta}_{ij} + \eta_{ij})/2} - \frac{{\rm d}u_j}{{\rm d}\delta_j}\Bigg\vert_{\delta_j = (\overline{\delta}_{ij} - \eta_{ij})/2} = 0\, ,
\end{equation}
which, when solved, determines the difference in cross-section displacements $\eta_{ij}$ at fixed curve separation.
While this procedure can be done for general monotonic functions $u_i$, for this work, we consider the case of simple power law forms 
\begin{equation}
    u_i(\delta_i) = c_i\delta_i^{p+1}
\end{equation}
with $c_i > 0$.
Solving Eq.~\ref{eq:contact_int_equilibrium_conditions}, we find
\begin{align}
    \eta_{ij}(s,s') &= \frac{c_j^{1/p} - c_{i}^{1/p}}{c_i^{1/p} + c_j^{1/p}}\left[R_i + R_j - \gamma_{ij}(s,s')\right] \\
    u_{ij}(s,s') &= \left(\frac{ c_i^{1/p}c_j^{1/p} }{c_i^{1/p} + c_j^{1/p}}\right)^p\left[R_i + R_j - \gamma_{ij}(s,s')\right]^{p+1}
\end{align}
allowing for simultaneous prediction of the point of contact between points $\bm{\gamma}_i(s)$ and $\bm{\gamma}_j(s')$ as well as the effective pairwise contact energy.
Note that in the case where $p = 3/2$, we recover the standard generalized form for Hertzian contact interactions~\cite{LandauLifshitz1986} and in the case where $c_i = c_j$, we recover $\eta_{ij} = 0$.
We finally determine the deformed radius of the curve at vertex $i$ to be
\begin{equation}
    \begin{split}
        R_i' &= R_i - \frac{R_i+R_j -\gamma_{ij}}{1 + \left(\frac{c_j}{c_i}\right)^{1/p}} 
    \end{split}
\end{equation}
where the equivalent expression for $R_j'$ is simply found by exchanging $i\leftrightarrow j$.

\subsection{Computational method}

We consider pairs of curves with cross-sectional radii $R_1$ and $R_2$ such that their sum $D = R_1 + R_2$ remains fixed at a value.
The energy can then be parametrized in terms of a dimensionless bending modulus $\tilde{B} \equiv B/(Y D^2)$ and dimensionless compressive stiffnesses $\tilde{c}_i \equiv c_i D^{p+3}/Y$.
For these computational studies, we take $D = 0.3$, $\tilde{B} \approx 1.11\times 10^{-2}$, and $\tilde{c_i} \approx 4.44\times 10^{-2}$, where $p = 3/2$ to model Hertzian-style interactions.
These parameters ensure that the curves are approximately inextensible and while the contact interactions are stiff enough to prevent topology-altering crossings under moderate stresses, they are still soft enough to permit significant deformation of the cross-section.
In order to minimize the total energy, we approximate the shape of the curves using the discrete elastic rod (DER) model advanced by Bergou, et al.~\cite{Bergou2008}, which faithfully preserves the continuum limit of the bending and stretching energies.
Meanwhile, the contact energy is computed through quadrature, with evaluations of the double integral in Eq.~\ref{eq:contact_int} accelerated through use of a k-d tree for efficient tracking of pairwise interactions.
Finally, we minimize the total energy through an implementation of the FIRE algorithm for gradient-based structural relaxation~\cite{Bitzek2006}.

\subsection{Construction of medial surfaces}

A Voronoi tessellation of space generated by point sets provides one way of defining the ``neighborhood'' around each particle~\cite{Lazar2022}.
Given a set of generating points $\left\{p_i\right\}$, the standard Voronoi tessellation creates a cell $\mathcal{C}_i \subseteq \mathbbm{R}^d$ containing a single generating point $p_i$, such that any arbitrary point $q$ belongs to cell $\mathcal{C}_i$ if $\lVert q-p_i\rVert < \lVert q-p_j\rVert$ for all $j \neq i$.
The interiors of each cell are disjoint, ${\rm int}(\mathcal{C}_i) \cap {\rm int}(\mathcal{C}_j) = \empty$ for $i\neq j$ so that the collection of Voronoi cells $\bigcup_i \mathcal{C}_i$ provides a tessellation of space.
The boundaries $\partial \mathcal{C}_i$ of Voronoi cells contain the collection of points that are equidistant to two or more generating points; that is, the locus of points $q$ satisfying the condition $\lVert q - p_i \rVert = \lVert q - p_j \rVert$ defines a shared boundary $\partial \mathcal{C}_i \cap \partial \mathcal{C}_j \neq \emptyset$.
Voronoi tessellations are found in contexts as diverse as granular packings~\cite{Behringer2019} and defects in crystal systems~\cite{Lazar2022,Feller2025} to urban planning~\cite{She2015}.

\begin{figure}
    \centering
    \includegraphics[width=\linewidth]{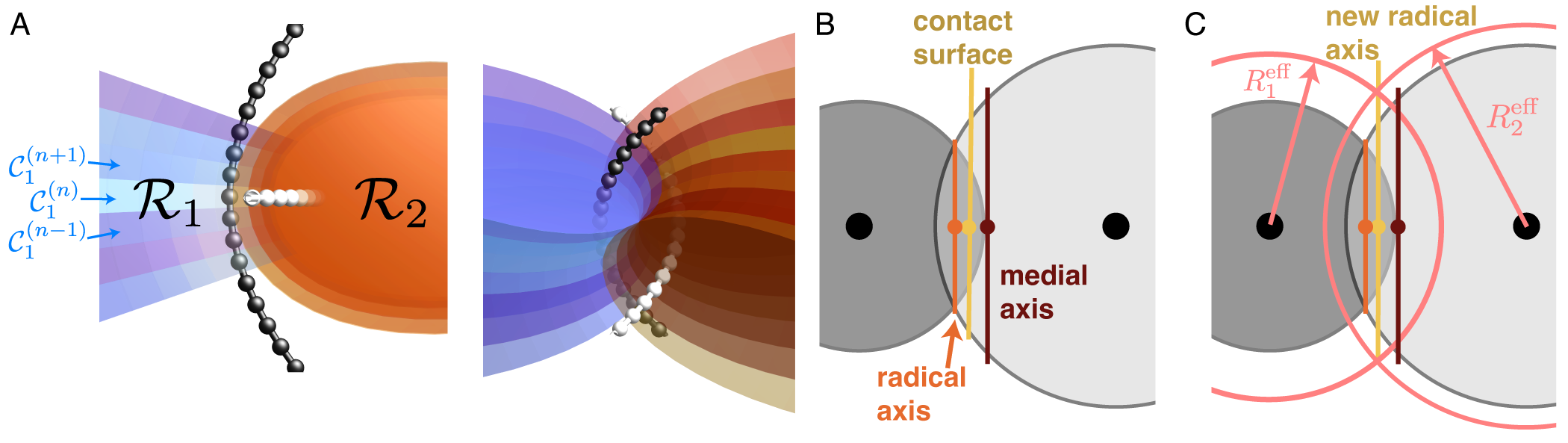}
    \caption{(A) Close-up of a pair of discretized curves, along with a collection of Voronoi cells $\mathcal{C}$ generated by curve vertices. The union of these cells associated with a given curve generates a medial region $\mathcal{R}$, the boundary of which is the medial axis transform. (B) A representation of the difference between the medial axis and the radical axis generated by pairs of disks of different radii, along with a distinct contact surface. (C) With a suitable transformation of disk radii, the radical axis can be translated to the contact surface.}
    \label{fig:2}
\end{figure}

The medial axis transform generalizes the construction of Voronoi tessellations to higher-dimensional generating sets (i.e.~rather than just generating points), resulting in much more complex partitions of space.
Originating as a method for shape recognition~\cite{Blum1973}, this technique has recently found use in the construction of interlocking tiles generated by curves~\cite{Akleman2020,Estrin2021,Krishnamurthy2022} as well as thermodynamics models of bicontinuous phases of block copolymers~\cite{reddy_medial_2022,dimitriyev_medial_2023,Dimitriyev2025}.
Instead of a convex polyhedral cell, the regions $\left\{\mathcal{R}_i \right\}$ generated by a set of curves $\left\{\bm{\gamma}_i\right\}$ are generally non-convex partitions of space that may not be simply connected: they generally contain handles.
In brief, a region $\mathcal{R}_i$ consists of all points $q$ such that $d(q,\bm{\gamma}_i) < d(q,\bm{\gamma}_j)$ for $i\neq j$, where the distance between a point and a generating curve is given by the infimum of the distance between $q$ and any point on the curve, i.e.~$d(q,\bm{\gamma}_i) = {\rm inf}_{s}\lVert q - \bm{\gamma}_i(s)\rVert$ where $s$ is the curve's arclength parameter.

One way of algorithmically constructing the medial axis transformation is through (i) discretization of the generating curves into a set of points, (ii) creation of a Voronoi tessellation generated by the discrete set of points, and finally (iii) identifying a region $\mathcal{R}_i$ as the union of constitutive Voronoi cells.
That is, a curve $\bm{\gamma}_i$ is discretized into a set of points $\left\{p^{(n)}_i\right\}$, where $p^{(m)}_i = \bm{\gamma}_i(s^{(m)})$ for an ordered collection of arclength parameters $s^{(m)} \in [0,L_i]$ (Fig.~\ref{fig:1}A).
Gathering all discrete points from each curve, we then generate a Voronoi tessellation of space, resulting in cells $\mathcal{C}^{(m)}_i$.
The medial region $\mathcal{R}_i$ containing generating curve $\bm{\gamma}_i$ is then approximated by the union of curve-associated Voronoi cells, $\mathcal{R}_i \approx \bigcup_n \mathcal{C}_i^{(m)}$, as depicted in Fig.~\ref{fig:2}A.
Algorithmically, this union is formed by constructing the set of cells $\mathcal{N}_i^{(m)}$ neighboring a specific cell $\mathcal{C}_i^{(m)}$, such that $\mathcal{C}_j^{(n)} \in \mathcal{N}^{(m)}_i$ if $\partial\mathcal{C}_i^{(m)} \cap \partial\mathcal{C}_j^{(n)} \neq \emptyset$.
We then remove the boundary between Voronoi cells whenever $i=j$.
As a result, $\mathcal{R}_i$ is a closed region of space enclosed by a polygonal meshed approximation of its boundary, $\partial\mathcal{R}_i$.
To construct the Voronoi tessellations for this work, we use \texttt{Voro++} software~\cite{Rycroft2009}.

\subsection{Radical medial surfaces}

For polydisperse particle packings, it is common to use weighted Voronoi tessellations to account for the different volumes that each particle is expected to occupy.
Of the weighted tessellations, the Laguerre-Voronoi or ``radical Voronoi'' tessellation is perhaps the most common.
The construction replaces the normal Euclidean distance $d(q,p_i)$ use for the Voronoi tessellation with the ``power'' $\Pi_i(q) = d^2(q,p_i) - R_i^2$, where $R_i$ is the radius of the circle around generating point $p_i$.
The cell $\mathcal{C}_i$ then corresponds to the collection of points $q$ such that $\Pi_i(q)<\Pi_j(q)$ for all $i\neq j$.
The consequence of this is that if the spheres around generating points $p_i$ and $p_j$ overlap then $\mathcal{C}_i$ and $\mathcal{C}_j$ share a cell boundary consisting of a flat face that cuts through the circular cusp formed by the intersection of two spheres, as shown in Fig.~\ref{fig:2}B; this plane is called the ``radical axis.''
The radical axis forms an approximate contact region between two spheres, although, as we shall show, this is not the same contact region as the formed by power-law contact interactions. 

In the following, we construct a version of the medial axis transform using radical Voronoi cells using the same procedure as outlined in the previous section.
However, this ``radical axis transform'' imagines that each generating curve $\bm{\gamma}_i(s)$ has a smoothly-varying radius $R_i(s)$; the ``power'' between a point $q$ and a generating curve $\bm{\gamma}_i$ is then taken to be $\Pi_i(q) \equiv \inf_s\left(\lVert q - \bm{\gamma}_i(s)\rVert^2 - R^2_i(s) \right)$.
The algorithm for discrete generation of the radical axis transform then follows by analogy.

Since the radical axes of two spherical bodies assumes that the spheres can interpenetrate without deformation, the radical axis transform lacks the details of contact mechanics.
However, as suggested in Fig.~\ref{fig:3}C, we can trick the radical axis transform to find the correct contact points by simply adjusting the weights of the radical Voronoi tessellation.
This is done by determining effective radii $R_i^{\rm eff}$ such that the radical axis formed by taking these new radii into account passes through a contact point.
For simplicity, we will consider the case where $c_i = c_j$ but $R_i \neq R_j$.
One viable choice is given by
\begin{equation}
    \begin{split}
        R_i^{\rm eff}(s,s') &= R_i + \frac{1}{2}\frac{R_j^2 - R_i^2 - (R_j - R_i)\gamma_{ij}(s,s')}{R_i + R_j} \\
        R_j^{\rm eff}(s,s') &= R_j - \frac{1}{2}\frac{R_j^2 - R_i^2 - (R_j - R_i)\gamma_{ij}(s,s')}{R_i + R_j}
    \end{split}
\end{equation}
where the effective radii depend on the arclength parameters of both curves.
In most cases, the effective radii are not uniquely defined for a single generating point; in this case, we choose the largest value $R^{\rm eff} - R$.

\section{Studies of an orthogonal clasp}

\begin{figure}
    \centering
    \includegraphics[width=\linewidth]{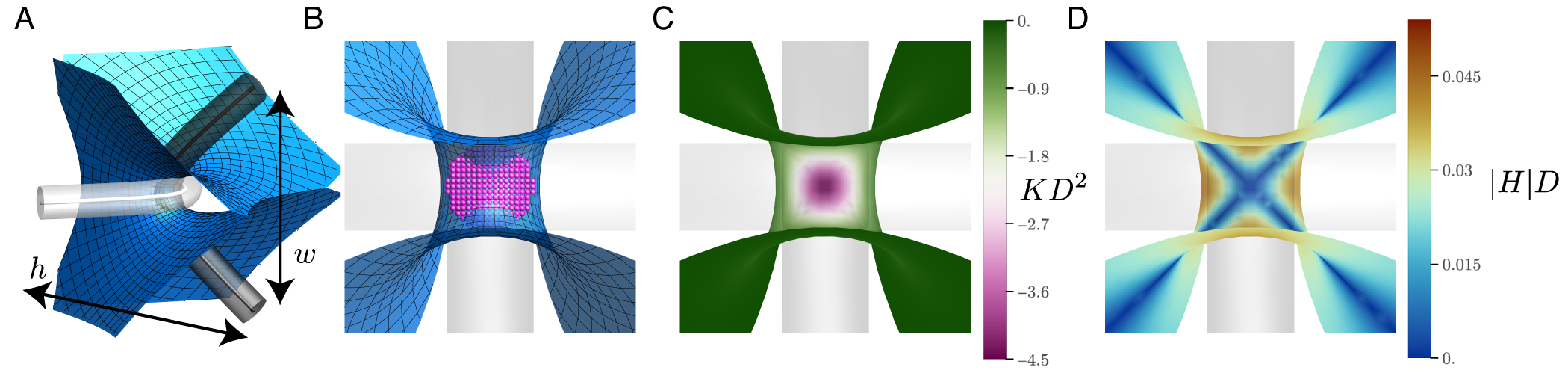}
    \caption{(A) Basic geometry of the symmetric orthogonal clasp, with a projection along a two-fold axis showing the distribution of contact points along the medial surface (B). (C) and (D) respectively show the Gaussian and mean curvatures, each non-dimensionalized by the effective fiber diameter $D$.}
    \label{fig:3}
\end{figure}

First, we consider the case of a clasp that consisting of a pair of identical curves oriented such that their tangent directions are orthogonal in the contact region (Fig.~\ref{fig:3}A).
Here, we select curves such that their rest lengths are $L/D \approx 9$, with endpoints of opposing curves spaced $h/D \approx 5.33$ from each other, and endpoints on the same curve spaced $w/D \approx 6.67$ apart.
This puts the two curves in contact with significant overlap, maintaining a stressed equilibrium configuration.
Plotting the distribution of contact points $\bm{\Gamma}_{12}(s,s')$ in Fig.~\ref{fig:3}B, the contact region has a diameter that extends nearly the entire diameter of a curve with a maximum penetration depth of $\delta/D \approx 0.25$.

In order to map the curvature of the medial surface, we start by transforming the polygonal mesh into a triangular mesh through the introduction of centroidal vertices.
The mean curvature $H$ and Gaussian curvature $K$ are computed using the ``cotan formula'' and angle deficit measurements, respectively, which are standard discrete approximations to surface curvatures evaluated at vertices (see e.g.~\cite{crane2018discrete}).
This approach leads to reasonably smooth distributions for both curvatures, shown in Fig.~\ref{fig:3}C-D.
The Gaussian curvature distribution shows large negative values in the vicinity of the contact points, indicating a local saddle structure that is easily seen in Fig.~\ref{fig:3}A.
Away from the contact region, the distribution rapidly approaches a flat or developable limit $K\to 0$.
The mean curvature distribution shows a richer texture, approaching zero at the center of the contact region and along quasi-1D paths that cross the contact region on a $45^\circ$ bias relative to the curve directions.
Interestingly, while the center of the contact region indeed approximates a patch of a minimal surface, there are large mean curvature gradients that develop slightly outside of the contact center, implying a strong deviations from minimal surface geometry.
On the other hand, the local mean curvature maxima in this region also correspond to regions where the contact points deviate from the medial surface, suggesting that the true contact surface may support a larger region of constant mean curvature than is suggested by the medial surface approximation.

\begin{figure}
    \centering
    \includegraphics[width=\linewidth]{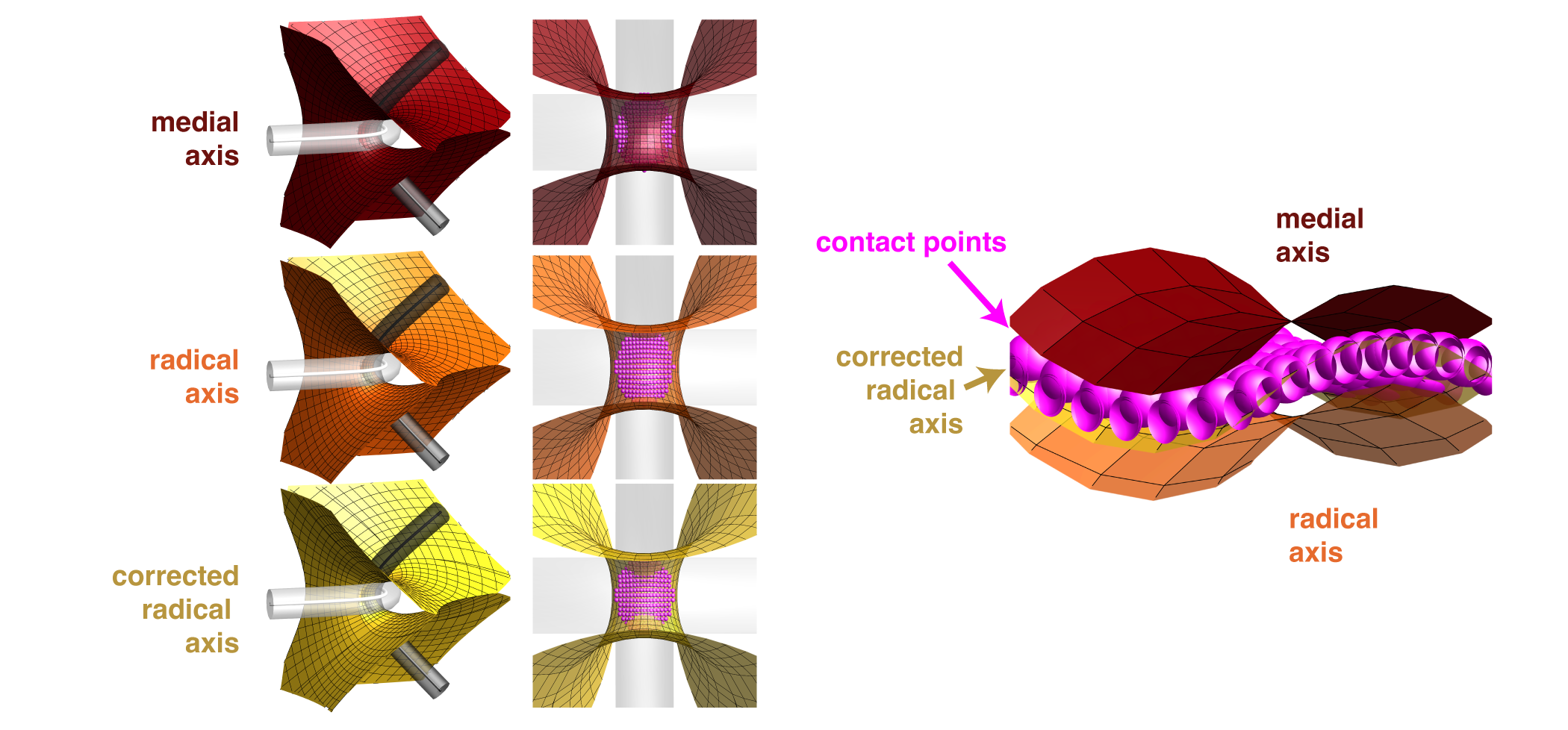}
    \caption{Medial and radical axes in the case of two curves of different radii. Magenta points represent the calculated contact points. The corrected radical axis provides the best approximation to the contact surface, as shown on the right.}
    \label{fig:4}
\end{figure}

\subsection{Asymmetric curves}

Next, we consider an example of an asymmetric clasp, where the two curves have different radii.
We choose $R_1 = 0.17$ and $R_2 = 0.13$ such that the effective diameter remains $D = 0.3$ while fixing the compressive stiffness parameter $c_1 = c_2$ as before.
As discussed in an earlier section, we expect that this asymmetry will deflect the contact surface from the medial surfaces generated by the centerlines.
In Fig.~\ref{fig:4}, we compare the medial surface against two other surfaces.
The first is the radical axis found from the union of radical planes generated by treating each curve vertex as a generating point of either radius $R_1$ or $R_2$.
The second is a ``corrected'' radical axis found by selecting $R_1^{\rm eff}$ and $R_2^{\rm eff}$ at each generating point such that the radical planes lie at the predicted contact points.
Note that this method isn't perfect: since each generating point on a curve is typically in contact with multiple points from the other curve, there can be many values for the corrected radius; of these possibilities, we always select the corrected radii that are furthest from the original radii.
While each of these surfaces are quite similar, a closer look reveals a slight stratification between the surfaces, where the corrected radical axis lies between the medial axis and the un-corrected radical axis.
Furthermore, it is clear that the corrected radical axis lies closer to the contact points and thus provides a better approximation to a contact surface.

\begin{figure}
    \centering
    \includegraphics[width=\linewidth]{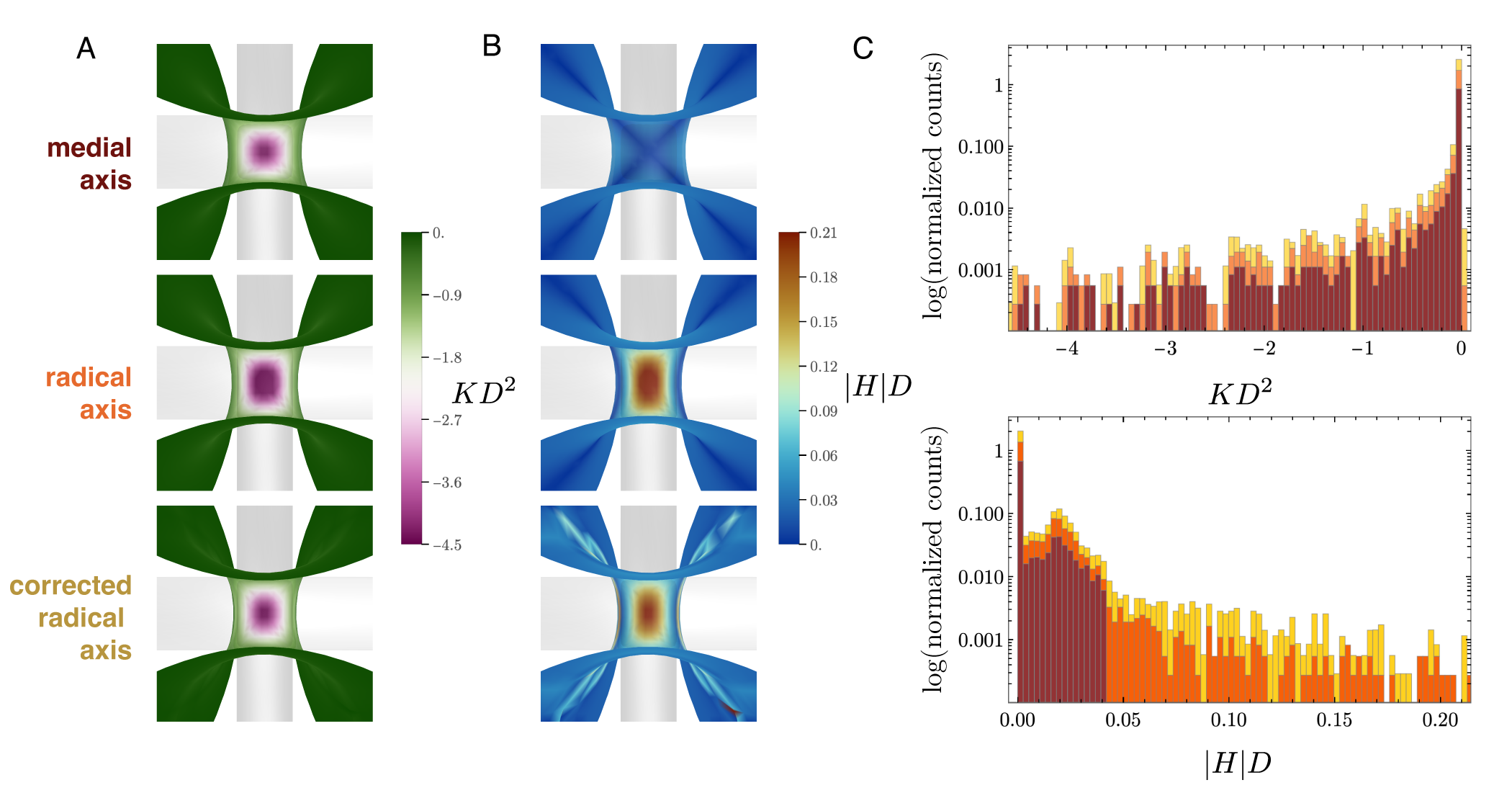}
    \caption{Comparisons of the Gaussian curvature (A) and mean curvature (B) distributions for three candidate approximations to the contact surface for an asymmetric clasp. Log-scale histograms of curvature data are shown in (C) and (D). Note that each data set is interdependently normalized and that the histograms are stacked.}
    \label{fig:5}
\end{figure}

What can we learn from this new construction? 
To answer this, we take a close look at the curvature distributions of each candidate surface in Fig.~\ref{fig:5}.
While the Gaussian curvature distributions shown in Fig.~\ref{fig:5}A appear similar for each surface, the mean curvature distributions shown in Fig.~\ref{fig:5}B reveal some interesting distinctions.
The medial axis has a much smaller magnitude of mean curvature than the two radical axes, with maxima differing by an order of magnitude.
The radical axes show at least a five-fold increase compared with the symmetric clasp case in Fig.~\ref{fig:5}D.
This can be attributed to the inherent asymmetry of a thinner curve wrapping a thicker curve: while the negative Gaussian curvature of these saddle-shaped surfaces tell us that the principal curvatures have opposite signs, the asymmetry of curve wrapping requires that the principal curvatures differ in magnitude, resulting in non-zero mean curvature.
Notably, the corrected radical axis has a slight reduction in the averaged mean curvature compared with the un-corrected radical axis, suggesting an energetic preference towards smaller mean curvature. 
However, this effect appears slight, as the (stacked) Gaussian curvature and mean curvature histograms (Fig.~\ref{fig:5}C) show little distinction, although it appears that the uncorrected radical axis has a longer high-H tail than the corrected radical axis.
This calls for a longer study over a wider parameter range, which falls outside the scope of this manuscript.

\section{Studies of a simple knit}

\begin{figure}
    \centering
    \includegraphics[width=\linewidth]{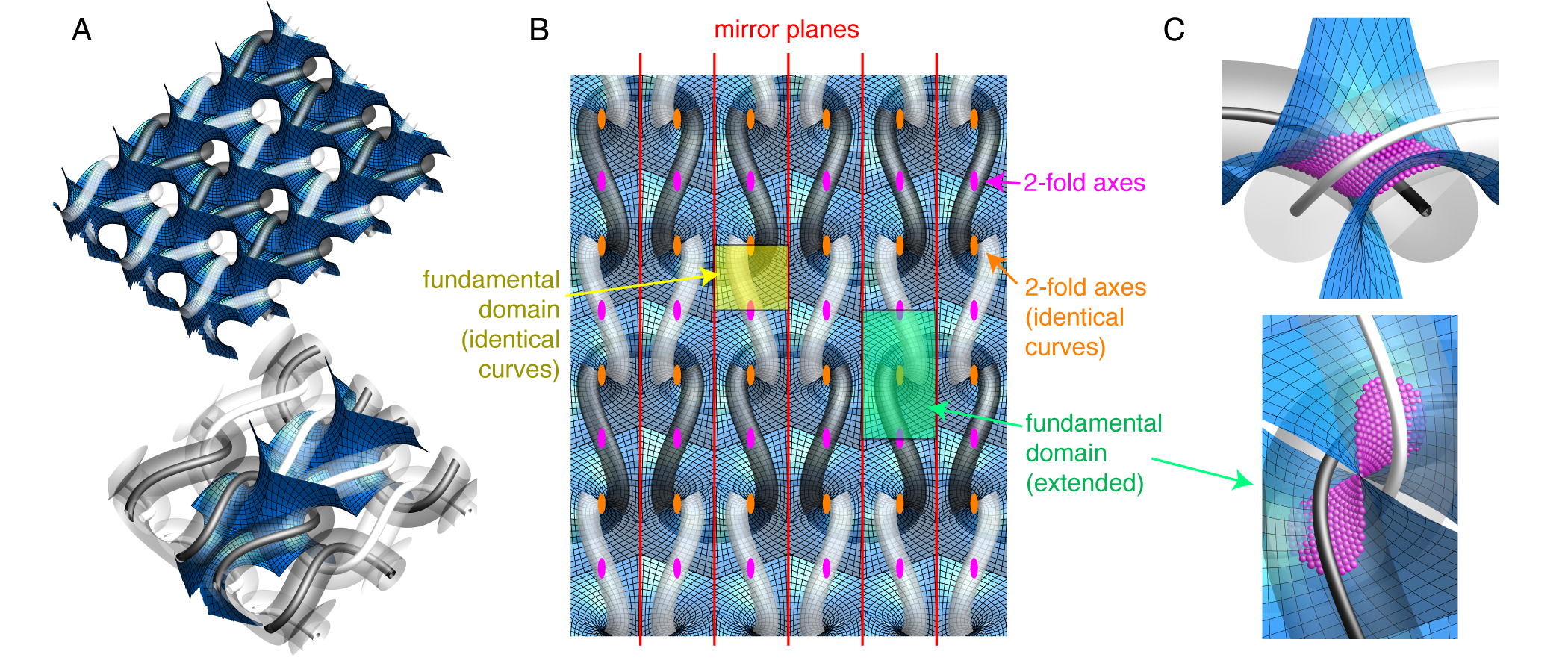}
    \caption{(A) View of a simple knit along with the medial axis transform generated in the case that both families of curves are identical. (B) Map of mirror and two-fold rotation symmetries in a simple knit. Note that half of the two-fold axes (orange) are removed in the case that the two families of curves are no longer identical; the magenta two-fold axes remain, however. In the case of asymmetry between curves, an extended fundamental domain (green) must be used instead of the usual fundamental domain (yellow). (C) Views of the extended fundamental domain showing contact points (magenta) in close proximity to the medial surface (blue).}
    \label{fig:6}
\end{figure}

While there is a variety of commonly used knitting patterns, we will restrict ourselves to the most basic, which consists of a single type of stitch that is repeated on a rectangular lattice.
This ``simple knit,'' shown with medial axes in Fig.~\ref{fig:6}A, is also known as stockinette or jersey and is by far the most common~\cite{singal_programming_2024,dresselhaus2026textilestwistedyarntopology}.
We will take advantage of the symmetries of the simple knit and reduce it to a representative structure in a fundamental domain.
Even though simple knits most often contain one type of yarn, we will go beyond this and consider knits consisting of two pairs of yarn.
Therefore, we choose a larger fundamental domain, shown in Fig.~\ref{fig:6}.
Despite the more complex, doubly-periodic structure, the stitch is still characterized by certain symmetries.
Principal among these is the mirror symmetry that splits the unit cell in half.
Furthermore, in the case the two curves are identical, there is a pair of two-fold rotational symmetries that help determine the boundary of the fundamental domain.
We therefore identify the symmetric simple knit as belonging to layer group \texttt{pma2} with orbifold notation $22\ast$~\cite{Mahmoudi2026}.
In the case where the two curves are distinct, one half of the two-fold rotational symmetries disappear, leaving behind a reduced set of symmetries.
However, the group structure is maintained with the definition of an expanded fundamental domain.

\subsection{Symmetric curves}

\begin{figure}
    \centering
    \includegraphics[width=\linewidth]{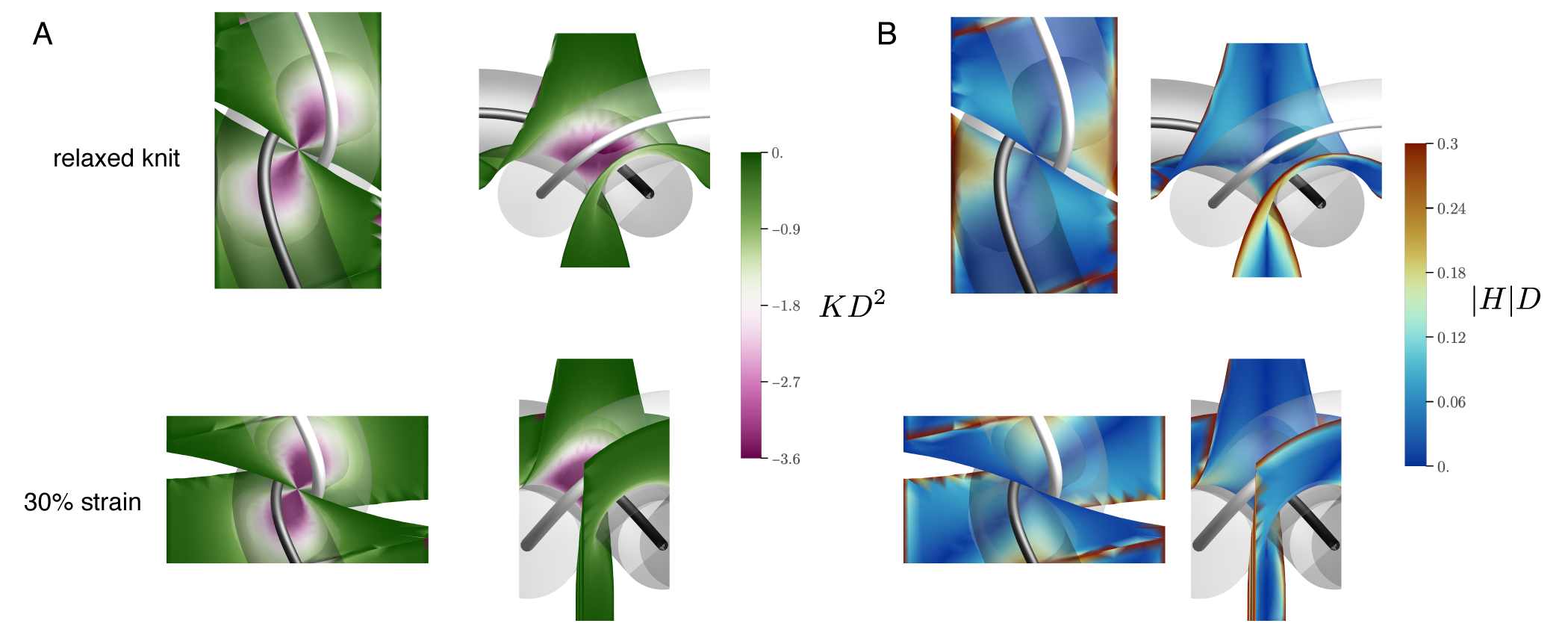}
    \caption{Gaussian curvature (A) and mean curvature (B) distributions for a relaxed knit and one stretched along the course direction by 30\%.}
    \label{fig:7}
\end{figure}

Mirroring the clasp study, we start with the fully symmetric, relaxed structure.
This relaxed stitch is found by allowing the dimensions of the unit cell to relax until the energy is minimized, resulting in cell dimensions $a/D \approx 3.18$ and $b/D \approx 5.49$.
As shown in Fig.~\ref{fig:6}A, the medial axis transform results in a periodic handlebody with patches that resemble the helicoid array scaffold found by other authors~\cite{Knittel2020,Wadekar2020_helicoidal,Wadekar2021,Niu2024}.
Unlike the helicoid array, the medial surface fully partitions space by deviating into flat patches that extend away perpendicular from the fabric's plane.
However, these flat patches are not significant features of the contact surfaces of the knitted fabric: as shown in Fig.~\ref{fig:6}C, the contact points lie on the helicoidal part of the medial surface.
This is confirmed by the large values of negative Gaussian curvature localized in regions containing the contact points, shown in Fig.~\ref{fig:7}.

Similar to the clasp, the mean curvature distribution exhibits considerable variation.
This variation is pronounced in the region where the yarn crosses the mirror symmetry plane.
We additionally stretch the knit by 30\% strain along the course direction and find little change in mean curvature but an increase in Gaussian curvature, as shown in Fig.~\ref{fig:7}.
This is likely a result of the preservation of the clasp's 2-fold rotation symmetry under this deformation.
The increase in the magnitude of the Gaussian curvature comes from an increasing concentration of fiber curvature in the clasp region, which has been found in previous mechanical studies of knitted fabrics~\cite{singal_programming_2024}.

\subsection{Asymmetric curves}

\begin{figure}
    \centering
    \includegraphics[width=\linewidth]{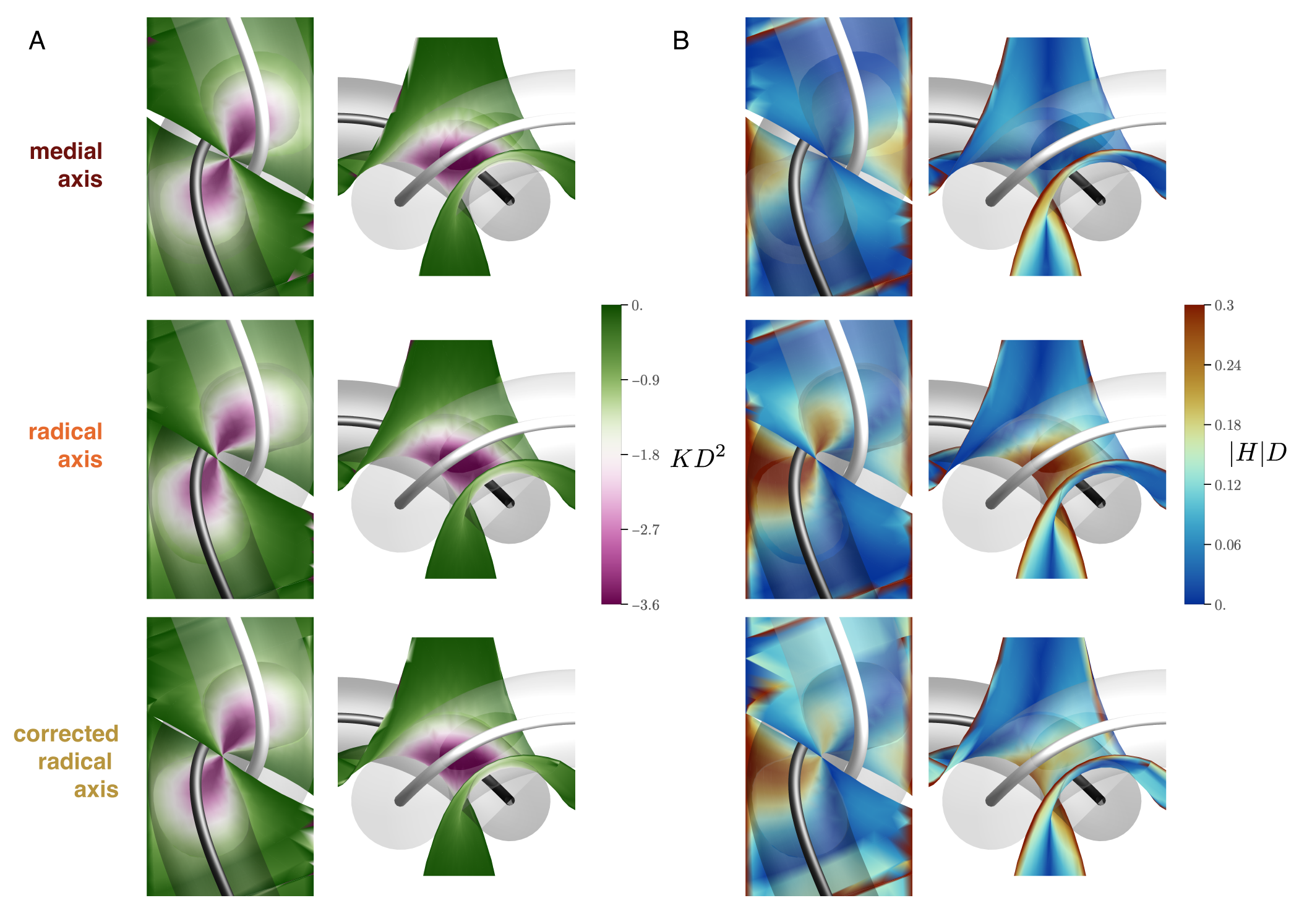}
    \caption{Gaussian curvature (A) and mean curvature (B) distributions for two families of curves with different radii, plotted over the fundamental domain.}
    \label{fig:8}
\end{figure}

Similar to the orthogonal clasp, we now consider a simple knit constructed from alternating rows of curves with different radii, $R_1 = 0.17$ and $R_2 = 0.13$.
Figure~\ref{fig:8} compares the curvature distributions for the different candidate approximations for contact surfaces: the medial axis, the un-corrected radical axis, and the correct radical axis.
All three have essentially the same pattern and scale of Gaussian curvature, which adopts large, negative values in the clasp-like regions before approaching zero further away.
The mean curvature distribution varies considerably but is similar to the clasp, the medial axis exhibits near-zero mean curvature throughout the region near curve contacts, only deviating further away.
The uncorrected radical axis shows the largest variation and attains a larger maximum mean curvature, whereas the corrected radical axis depresses this value somewhat and lies between both extremes.
This further supports a mechanism that reduces the mean curvature for contact surfaces.

\section{Conclusion}

In this work, we have revisited a basic model for contact interactions between curves and showed that we can extract new information, namely an approximation for the contact surface.
We can do this even without a solid mechanics or 3D continuum elasticity description of the interacting curves.
Moreover we addressed the hypothesis that the contact surfaces are well-approximated by minimal surfaces and found that near-minimal surfaces can be found in certain cases of high symmetry, but deviations from constant mean curvature develop rapidly as one looks around the neighborhood of a contact region.
In order to handle asymmetry between curves, for our cases due to differences in yarn radii, we introduced an extension of the medial axis transformation based on radical Voronoi tessellations.
This new method, along with a correction that allows variations in an ``effective'' curve radius, creates reasonable approximations to contact surfaces; these ``corrected radical axes'' perform better than the standard Voronoi-based approach. 
We applied this method to both an orthogonal clasp and a simple knit structure.
Our results open the way for more in-depth studies of how contact surfaces change in shape in response to either asymmetry in the curves themselves or the influence of external loading. 
In particular, it would be useful to validate this approach against experiments and finite element analysis.
There also remains an open question of how changing the constitutive model alters our results; the Hookean form, while commonly used in modeling, is otherwise poorly motivated for curves and instead there are empirical models that have been fitted to data (see e.g.~\cite{singal_programming_2024}).

\subsubsection{Acknowledgments} 
We have benefited from stimulating conversations with Lauren Niu, Randall Kamien, Vinayak Krishnamurthy, Ergun Akleman, and Vanessa Sanchez. We furthermore thank Sonia Mahmoudi, Samuel Poincloux, and Tohoku University for hosting one of us (MSD) for the ``Mathematics \& Physics of Entangled Metamaterials'' workshop, which helped initiate key elements of this work.

\bibliographystyle{splncs04}
\bibliography{refs}

\end{document}